# Random Forest Classifier Based Prediction of Rogue waves on Deep Oceans

Pujan Pokhrel
Canizaro Livingston Center for Gulf Informatics
*University of New Orleans*
New Orleans, LA, United States
ppokhrel@uno.edu

Elias Ioup
Naval Research Laboratory
*Stennis Space Center*
Missisipi, United States
elias.ioup@nrlssc.navy.mil

Md Tamjidul Hoque[†]
Canizaro Livingston Center for Gulf Informatics
University of New Orleans
New Orleans, LA, United States
thoque@uno.edu

Julian Simeonov
Naval Research Laboratory
Stennis Space Center
Missisipi, United States
julian.simeonov@nrlssc.navy.mil

Mahdi Abdelguerfi
Canizaro Livingston Center for Gulf Informatics
University of New Orleans
New Orleans, LA, United States
mahdi@cs.uno.edu

[†]*To whom correspondence should be addressed*

## Abstract

*In this paper, we present a novel approach for the prediction of rogue waves in oceans using statistical machine learning methods. Since the ocean is composed of many wave systems, the change from a bimodal or multimodal directional distribution to unimodal one is taken as the warning criteria. Likewise, we explore various features that help in predicting rogue waves. The analysis of the results shows that the Spectral features are significant in predicting rogue waves. We find that nonlinear classifiers have better prediction accuracy than the linear ones. Finally, we propose a Random Forest Classifier based algorithm to predict rogue waves in oceanic conditions. The proposed algorithm has an Overall Accuracy of 89.57% to 91.81%, and the Balanced Accuracy varies between 79.41% to 89.03% depending on the forecast time window. Moreover, due to the model-free nature of the evaluation criteria and interdisciplinary characteristics of the approach, similar studies may be motivated in other nonlinear dispersive media, such as nonlinear optics, plasma, and solids, governed by similar equations, which will allow for the early detection of extreme waves.*

## 1. INTRODUCTION

Rogue waves are classified as waves having a height more than 2.2 times the significant wave height (Hs) in the wave field [1]. Sometimes, they are studied using various nonlinear equations, which assume that wave energy gets focused on these events and generates nonlinearity. In this paper, rogue waves have been studied under the second assumption. Rogue waves are observed in hydrodynamics [1], optics [2], quantum mechanics [3], Bose-Einstein condensates [4], and finance [5]. They are mainly studied analytically using the spectral algorithms applying some deterministic equations like the nonlinear Schrodinger equation [6, 7]. Although rogue waves may be needed in fiber optics to satisfy certain energy levels and to locate the information using matched filtering, they are dangerous in oceans and present a danger to the safety of marine operations. Examples of these events include the sinking of Prestige [8], El Faro [9], and damage to the Draupner platform [10]. To prevent these accidents, an early detection system with precise emergence time of these events is needed.

There are various methods for the early detection of nonlinear waves. For example, spectral techniques can be used by measuring the super-continuum patterns in the Fourier spectra before the rogue waves form [11]. However, checking the Fourier spectra solely would fail to give any clue about the expected emergence point (or time) of a rogue wave in a chaotic wave field. Although various spectral methods have also been proposed to include the time-dependent information about the waves [12, 13], the prediction time is only in the order of seconds. While such short time scales may be beneficial for saving lives in oceans, they are not enough for avoiding exposure to such events. Later, Birkholz *et. al.* [14] proposed a Grassberger-Procaccia nonlinear time series algorithm for the prediction of rogue waves and have slightly improved the time scale. Likewise, AD Cattrell [15] proposed that machine learning/statistical methods could be used to predict rogue waves using characteristic wave parameters.

To achieve the goal of forecasting rogue waves, it is necessary to develop statistics based computational approaches that can reliably and rapidly identify and forecast rogue waves in chaotic wave fields like the oceans. In contrast with the deterministic equations, such statistical methods can be employed for predicting a wide range of instabilities and can also help simulate the physics of the equations without computing a set of equations periodically. Some of the classical nonlinear evolution equations include nonlinear Schrodinger equation [6, 7, 16], Davey-Stewartson system [17, 18], Korteweg-de Vries equation [19], Kadomtsev-Petviashvili equation [20], Zakharov equation [21] and fully nonlinear potential systems [22]. However, such equations only describe a specific instability and using a set of equations every time to forecast rogue waves is not possible for a continental/planetary scale prediction. Since the ocean waves are often bimodal/multimodal due to the presence of many wave-systems, it is assumed that rogue waves are more likely to occur when the distribution turns unimodal. Afterwards, we used various statistical machine learning methods to forecast rogue waves [23-27].

## 2. POSSIBLE CAUSES OF ROGUE WAVES

Various methods for the formation of rogue waves have been explored in the literature. Some of them include (a) Linear Superposition, (b) Nonlinear effects, and (c) wind-wave interactions.

### a) Linear Superposition and weakly nonlinear effects

The most widely used theory for describing statistics of the surface gravity waves is the Gaussian theory, which assumes that waves are a linear phenomenon. However, the theory fails to account for nonlinear effects [28-32]. Considering the waves are weakly nonlinear, various methods have been developed for the narrowband unidirectional case [6, 28, 30, 32-38]. A more general theory for second-order interactions of waves in the random directional sea was derived by Sharma and Dean [39], which in theory should be able to capture the effects of wave steepness, water depth and directional spreading with no approximations other than the truncation of the small amplitude expansion to the second order. However, this model only attributes the abnormal events in the sea to the linear superposition of the waves in the oceans. It is important to note that at the second-order in Stokes expansion, the crests are sharper and higher while the troughs are flatter and lower; there is also a long wave set down (e.g., see [40]), which causes a decrease in the mean surface in regions of groups of high waves. Supported by in-situ measurements, Forristall [41] looked at the distribution of simulated second-order long-crested (i.e. unidirectional) and short-crested waves and verified the predictions of the second-order theory. Linear superposition of waves thus remains one of the most likely mechanisms behind the formation of rogue waves. **Figure (1)** and **Figure (2)** show the evolution of the spectra when superposition is taking place. In such cases, more than one waves with similar frequency components and small incidence angles can combine to form a large wave.

Three mechanisms have been proposed to explain how superposition occurs. First, the waves of different scales and frequencies propagate at different speeds. Likewise, waves of the same scale propagate with different speeds depending on their steepness. Since the ocean wave spectrum is continuous, all the waves within the spectrum are present and propagate at the same time in a random wave field [40]. Similarly, they can intersect and pile-up resulting in a higher surface elevation. Likewise, the wave fields with the same frequency and same steepness can be focused and superposed if they come from different angles [42]. This phenomenon is also known as wave focusing. While focusing is linear in this case, the last stage of the focused-wave dynamics demonstrates various nonlinear behaviors when the steepness is large enough [43]. Wave focusing due to directionality has been found to be a regular cause for wave breaking in wave tanks, which is associated with large waves [22]. If the waves of the same scale come from different directions, then a superposition of only two waves is needed to double wave height and steepness. These conditions can produce regular events with the height being the summation of two wave heights [44] or, at certain angles, activate some mechanisms of wave instability [45, 46]. Linear superposition of waves is most likely at small angles (which is not too dangerous) or at angles close to 180 degrees, which have been shown to be dangerous even at low significant wave heights [44]. Thus, linear superposition remains one of the most likely mechanisms behind the formation of large waves in the oceans.

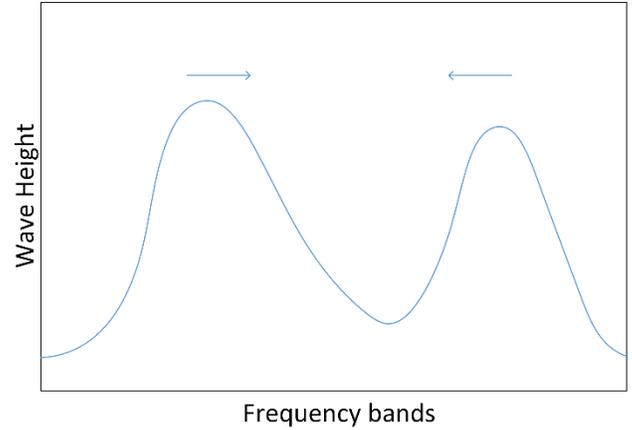

**Figure 1:** Example wave system with similar frequency components moving towards each other. Note that for superposition to occur, two waves have to move towards each other.

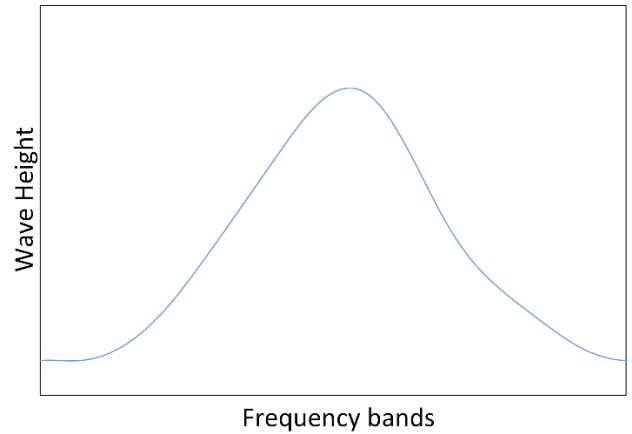

**Figure 2:** Superposition of two waves from Figure 1 to form a large wave. It shows how two waves traveling towards each other in Figure 1 can combine to form a large wave.

b)  **Nonlinear effects**

A thorough description of different aspects of abnormal waves has been provided by Kharif et al. [1]. The authors present various possible causes behind rogue waves like wave focusing and higher-order nonlinearities. One of the most studied higher-order instability in wave systems is the Benjamin-Feir instability due to third-order quasi-resonant interactions between the free waves when the initial spectra represent narrowband long-crested conditions [46-49]. Higher-order models have also been explored, which point to the fact that large waves may be caused by various nonlinear effects in the ocean[6, 7, 16, 22, 36, 50-52]. The likelihood of this mechanism is quantified by the Benjamin-Feir Index (BFI) [53] (see also [54] ). Favorable conditions for the instability can be generated mechanically in wave tanks [55-58] or simulated numerically [49, 52, 54, 59, 60]. Miguel Onorato [55] provided the first experimental evidence that nonlinear wave statistics, mainly in the wave tanks and shallow water conditions, depend on BFI. Likewise, from the results of Petrova and Guedes Soares [38, 61-63], it is known that, in general, the wave nonlinearity increases with the distance from the wavemaker on experiments on the wave tank. Numerical studies [49, 57] analyzing the effect of the directionality show that the wave trains become increasingly unstable towards long-crested conditions. However, the initial requirements for the instability make this mechanism unlikely to be the primary cause for most extreme wave occurrences in oceanic conditions, characterized by the broader spectra and directional spread [41, 61]. It is important to note that the nonlinear statistics of the following sea states observed were usually lower than the mixed crossing seas with identical initial spectra. The results for the distribution of the wave heights corroborate the conclusion of Rodriguez [64] that the existence of two wave systems of different dominant frequencies but similar energy contents result in the reduction of probability of wave height higher than the mean and the effect becomes more significant as the intermodal distance increases. The higher-order wave nonlinearity is reported to increase significantly with the observed probability of occurrence of large wave events [59]. The third order wave-wave interactions expressed quantitatively by the mean of the coefficient of kurtosis, are considered in the case of Benjamin-Feir Instability, regarded as quasi-resonant four-wave interactions [48, 65]. It is observed that the high-frequency spectral counterpart for both following and crossing seas show a decrease in peak magnitude and downshift of the peak with the distance, as well as a reduction of spectral tail when modulational instability takes place [38, 47, 49]. The authors stress that that the results from removing the second and third-order bound wave effects from the nonlinear surface profiles show that, in some cases away from the wave generator, the wave parameters of the non-skewed profiles continue deviating largely from the linear predictions which justifies the need of using higher-order models for the description of wave data when free wave interactions become relevant. The result is well confirmed by a recent numerical experiment by Manolidis et al [66]. **Figure (3)** and **Figure (4)** show the evolution of the spectra when modulational instability and higher order nonlinear effects are taking place.

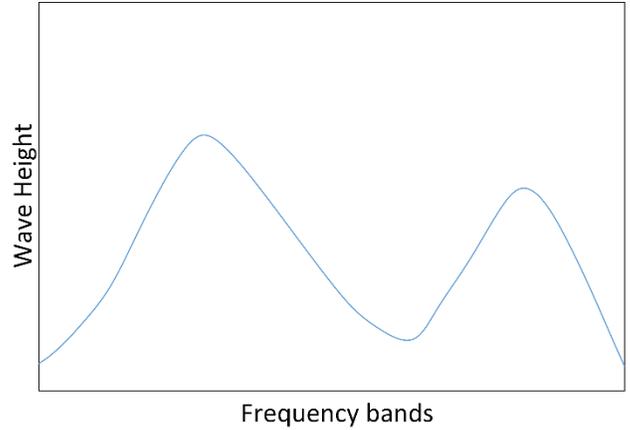

**Figure 3:** A directional distribution with two wave systems. The directionality of the waves is not essential for modulational instability.

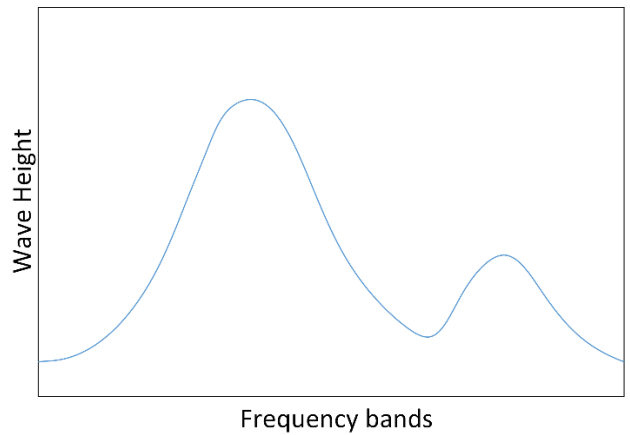

**Figure 4:** Example directional distribution when an extreme wave is occurring due to directional spreading. Note that as one wave undergoes the broadening of spectra, the other wave sucks up the energy from it to grow large.

c) **Wind-wave interactions**

It is known that storm waves are one of the five key processes identified in UKCP Marine Report that pose a great coastal risk in terms of flooding and inundation effects [67]. During storms, locally generated wind waves combine with the long period ocean swells to produce bimodal waves. Some reviews and comparisons of wave height distribution occurring after the storms in both deep and shallow water have been studied widely [30, 37, 41, 62, 68]. The analysis of the oceanic data collected in the stormy seas seems to indicate the validity of linear models for the distributions of large wave heights [30, 69, 70]. However, deviations between the theoretical predictions and observations occur at low probability levels when the measurements contain rare, huge waves, referred to as abnormal, rogue or freak waves [38].

Likewise, interactions between oceanic swell and sea components can cause nonlinear interactions required for freak waves to occur. The effect of combined sea components on the wave crest statistics, surface elevation, skewness, and kurtosis were shown for the first time by Bitner-Gregersen and Hagen [71] for the second-order time-domain simulations. Higher wave crests and larger nonlinear statistics have been reported for wind-dominated seas. Arena and Guedes Soares [37] performed Monte Carlo simulations of the second-order waves with the bimodal spectra representative of the Atlantic Ocean. They reported good statistical agreements between the empirical wave height distributions and the linear model of Bocotti [72-74] and also between the distributions of nonlinear wave crests/troughs and second-order formulation of Fedele and Arena [35]. Petrova [38] presented the results of the contribution of the third-order nonlinearity to the wave statistics both in terms of angle of incidence between the two crossing wave systems and the evolution of the waves along with the tank. The authors also conclude that distributions of the wave crests and troughs for a large angle of crossing seas are more likely to be predicted by weakly nonlinear models rather than the linear ones.

In some studies, such as Burcharth, Hawkes *et. al.*, Battjes, Reeve, [75-78] the phenomenon of wave bimodality has been elaborately described. Wind waves are characterized by one spectral peak with one significant wave height and one peak period. A bimodal (double-peaked) spectrum is usually formed through the combination of swell from a distant storm and locally generated wind sea. Transformations of these wave systems can be described in terms of wave crests, troughs and wave height distributions. Longuet-Higgins [79] proposed the Rayleigh distribution of wave heights and several modifications have been made to the low wave-height exceedance distributions [30, 62, 68, 72, 80-82]. Specifically, a depth modified version of the Rayleigh distribution was proposed by Battjes and Groenendijk [77] which is applied only to the unimodal waves. Similarly, Rodriguez [64] studied the wave height probability distributions using extracted gaussian bimodal waves from numerical simulations. The study classifies bimodal seas as wind-dominated, swell-dominated and mixed-sea conditions. Likewise, Petrova and Guedes Soares applied a linear quasi-deterministic theory to compare the energies from wind and swell seas using a simplified Sea-Swell energy ratio (SSER) on the assumption of wave nonlinearity [61]. Similarly, Norgaard and Lykke Anderson developed a slope dependent version of Rayleigh distribution based on an Ursell number criterion [83]. It should be noted that none of these studies have applied bimodal sea states that have varying proportions of swell, while at the same time containing a fixed amount of energy to investigate wave height distribution in shallow/deep water close to a structure, which is what occurs very often in practice. However, the studies are sufficient to conclude that the interactions of various swell-sea and wind-sea components can lead to the formation of a large wave.

## 3. RELATED METHODS

To the best of our knowledge, there exist two works for prediction of rogue waves: (a) work by Will Cousins *et. al.* [84, 85] and (b) work done at ECMWF by Janssen [86]. The work by Cousins *et al* focuses on the short-term prediction of extreme events in irregular unidirectional fields. The algorithm can only predict 2-3 minutes into the future and needs high-resolution information from LIDAR about all neighboring waves to make predictions. Thus, it can't be used for prediction for a longer time period. It is also essential to note that the waves in the oceans are rarely unidirectional. They are mostly bidirectional or multidirectional. Moreover, only Benjamin-Feir instability analyzed in the paper using nonlinear Schrodinger and other nonlinear equations to predict rogue waves. In the ocean fields, where it is common to have a bidirectional wavefield with a swell-sea component, the initial requirements for Benjamin-Feir instability to occur might not occur. Likewise, Janssen [86] proposed a shallow water version of the Freak wave warning system. The system is based on an estimate of kurtosis and skewness proposed for a narrowband version of the theoretical expressions for skewness and kurtosis. However, their system is only applicable to the shallow waters and narrow banded wave trains.

Thus, in this paper, we propose a novel method to predict rogue waves in oceanic waters under broader conditions. The only assumption we make in this paper is that since the ocean is bimodal or multimodal when the distribution turns unimodal, freak waves are more likely to occur. However, we do not make any assumption on the initial conditions for the nonlinear phenomenon to occur. This model-free assumption allows us to capture various nonlinear effects without being restricted by a specific equation. Moreover, we take the time window for each sample to be 26.67 min which is somewhat stationary. Due to the model-free nature of our evaluation mechanism compared to other methods, we propose that it should be able to capture various nonlinearities in the oceanic wavefields.

# 4. Development of Computational Methods for Forecasting Rogue Waves

## 4.1. Dataset

*The historical dataset of the oceanic waves around the United States is collected via CDIP buoys and is available at the NOAA website. We used the data from January 2007 to October 2019 for the study. Due to the huge amount of data, we only retained 40% of the total data as the benchmark dataset for the study. There are 754490 positive points and 189345 negative points in the benchmark dataset. Please note that the data was shuffled before the training and the testing phases to avoid any bias arising due to data collection.*

## 4.2. Feature selection for prediction of rogue waves

Many features have been used to identify rogue waves. The features used in this paper were derived from the Fourier spectra of the Directional Spreading Function after Discrete Fast Fourier Transform. It is important to note that the buoys don't always measure the same frequency components. It is thus necessary to derive the features that are adaptable to varying frequency components length. Likewise, these features also help to reduce the number of features significantly. The following features were derived from the four Fourier moments.

$$\sum_{i}^{n-1} mean(\sum_{j=i+i}^{n} minowski(X[:i], X[:j], k)) \quad (1)$$

$$\sum_{i}^{n-1} mean\_c(\sum_{j=i+1}^{n} mean\_c(X[i:j])) \quad (2)$$

$$\sum_{i}^{n-1} mean\_c(\sum_{j=i+1}^{n} median\_c(X[i:j])) \quad (3)$$

$$\sum_{i}^{n-1} mean\_c(\sum_{j=i+1}^{n} median\_c(X\_R[i:j])) \quad (4)$$

$$\sum_{i}^{n-1} mean\_c(\sum_{j=i+1}^{n} mean\_c(X\_R[i:j])) \quad (5)$$

$$\frac{1}{n}\sum kurtosis \quad (6)$$

$$\frac{1}{n}\sum skewness \quad (7)$$

$$\frac{1}{n}\sum energy \quad (8)$$

$$\frac{1}{n}\sum direction \quad (9)$$

where $mean\_c$ and $median\_c$ refer to column-wise mean and median respectively of the array derived from Fourier coefficients at different frequencies. Likewise, $X\_R$ refers to the array reversed in order, $minowski$ refers to the Minowski distance function. It is used because it is the generalization of Euclidean and Manhattan distances. The values of k used are 0 and 1. The value of $i$ and $j$ vary from 0 to 27 and cover the frequency bands in the spectra from 0.025 Hz to 0.580 Hz. Here, in equation (1) we calculate the mean Minowski distances between various frequency components. Measuring the mean distance will help us measure the intermodal distances, which in turn, helps identify various nonlinear effects. Likewise, the other equations (2), (3), (4) and (5) capture the information about the general shape of the directional distribution. It is done by measuring the average mean and median which helps identify skewed distributions. The main intuition behind the features is that rogue waves occur due to various linear/nonlinear interactions between waves and thus measuring how the mean and median fluctuates between different frequency components interact should capture more information about the rogue waves. Moreover, it is to be noted that when modulational instability occurs, it is characterized by the spreading of the initial narrowband spectra. In such cases when the wave-wave interactions occur, it is common to observe that one wave system grows at the expense of another. Thus, these features should be helpful in identifying various nonlinearities in the oceans.

Equation (6) and Equation (7) define the normalized kurtosis and skewness of the wave directional distribution. The equations for calculating skewness and kurtosis are described in section 4.3. Specifically, equations (14) and (15) describe formula to calculate skewness and kurtosis respectively. Likewise, equation (8) refers to the normalized sum of energy and equation (9) refers to the normalized sum of directions of different frequency components of Fourier spectra.

The other features derived are significant wave height, peak frequency, peak bandwidth, peak direction, total energy of the system and the dominant wave period.

## 4.3. Distinguishing unimodal sea state from bimodal/multimodal sea state

To calculate the unimodal sea state, we calculate the kurtosis and skewness from the Discrete FFT derived from the time series data at each frequency band. The assumption is that when the energy gets focused, nonlinear effects occur generating unimodal distribution and increasing likelihood for rogue waves to occur.

The estimate of the KVH criteria [87] is based on an integration over the frequency band 0.025 Hz to 0.580 Hz on the bulk Fourier moments a1, b1, a2, b2 weighted by the energy density.

$$\begin{bmatrix} a_1 \\ b_1 \\ a_2 \\ b_2 \end{bmatrix} = \frac{1}{E^b} \int_{0.025}^{0.580} (df\, E(f) \begin{bmatrix} a_1(f) \\ b_1(f) \\ a_2(f) \\ b_2(f) \end{bmatrix}) \quad (10)$$

where $E^b$ is the variance with

$$E^b = \int_{0.025}^{0.580} df\, E(f) \quad (11)$$

afterwards, we calculate

$$theta = tan^{-1}(\frac{b_1}{a_1}) \quad (12)$$

$$m_1 = (a_1^2 + b_1^2)^{1/2} \quad (13)$$

$$m_2 = a_2 * \cos(2 * theta) + b_2 * \sin(2 * theta) \quad (14)$$

$$n_2 = b_2 \cos(2\alpha) - a_2 \sin(2\alpha) \quad (15)$$

$$skew = (\frac{-n_2}{(1-m_2)})^{3/2} \quad (16)$$

$$kurtosis = \frac{(6 - 8m_1 + 2m_2)^2}{2(1-m_1)} \quad (17)$$

In equation (10), the bulk Fourier moments are derived to calculate the bulk KVH criteria. We take the integration of the Fourier moments multiplied by energy and bandwidth. It is then normalized by dividing it with variance calculate in equation (8). Likewise, equation (10) is used to calculate the bulk Fourier moments. Afterward, we use equation (12), (13) and (14) to find different parameters which are used to calculate skewness and kurtosis. Finally, we calculate the skewness and kurtosis in equation (16) and (17), respectively. Afterward, the KVH criteria are used to determine the unimodal distribution. KVH showed that the skewness and kurtosis are very sensitive to the secondary directional peaks and thus can be used to identify bimodal/multimodal distribution from a unimodal one. Although the KVH criteria are derived on the assumption of the unimodal distribution and describe the two peaked spectra as a warning criterion, we use the same criteria because of its model-free attributes but define the warning criteria as to when a unimodal distribution arises. The KVH criteria are given in the equations (18) and (19).

$$kurtosis < 2 + |skew|\, and\, |skew| \leq 4 \quad (18)$$
$$kurtosis < 6\, and\, |skew| > 4 \quad (19)$$

## 5. RESULTS AND DISCUSSIONS

In this section, we present the results of the experiments that were carried out in this study. All the methods were implemented using python language. The Scikit-learn library [88] was used for implementing the machine learning algorithms. Note that 10 fold cross-validation was used for testing the classifiers. Please note that the window of 26.67 min corresponds to 1600s and is a standard for most NOAA buoys.

### a) Performance of Logistic Regression with and without the spectra features

In this section, we compare the performance of the Logistic Regression classifier with and without the features (1) to (9). The classifier that didn't include the novel features contained significant wave height, peak frequency, peak bandwidth, the total energy of the system and the dominant wave period as features. Please note that the window of 26.67 min corresponds to 1600s and is a standard for most NOAA buoys.

*Table 1: Performance of Logistic Regression with and without Spectral features*

| Methods | With | Without |
|---|---|---|
| Sensitivity | **0.9125** | 0.7650 |
| Specificity | **0.7570** | 0.4520 |
| Balanced Accuracy | **0.8347** | 0.6085 |
| Overall Accuracy | **0.8663** | 0.4785 |
| FPR | **0.2429** | 0.5481 |
| FNR | **0.0874** | 0.2350 |
| Precision | **0.8988** | 0.1150 |
| F1 | **0.9056** | 0.1990 |
| MCC | **0.6768** | 0.1220 |

Value in **bold** indicates the best outcome.

As we can see from **Table 1**, the performance of the Logistic Regression classifier increases when we use the spectral features proposed in equations (1) to (7). We thus take the features and test various machine learning algorithms to choose the best model. Note that the same features have been used for predicting rogue waves for all the time windows.

### b) Search for the best classifier for the time window 0-26.67 min

*Table 2: Search for the best classifier for prediction*

| Methods | LogReg | KNN | RF | ET |
|---|---|---|---|---|
| Sensitivity | 0.9125 | 0.9233 | **0.9474** | 0.9339 |
| Specificity | 0.7570 | 0.7259 | 0.8332 | **0.8405** |
| Balanced Accuracy | 0.8347 | 0.8246 | **0.8903** | 0.8872 |
| Overall Accuracy | 0.8663 | 0.8646 | **0.9135** | 0.9061 |
| FPR | 0.2429 | 0.2741 | 0.1667 | **0.1594** |
| FNR | 0.0874 | 0.0766 | **0.0526** | 0.0660 |
| Precision | 0.8988 | 0.8885 | **0.9308** | 0.9327 |
| F1 | 0.9056 | 0.9055 | **0.9390** | 0.9333 |
| MCC | 0.6768 | 0.6685 | **0.7908** | 0.7751 |

LogReg=Logistic Regression, KNN=K Nearest Neighbors, RF=Random Forest, ET = Extra Tree

From **Table 2**, we can see that Random Forest performs the best with a Sensitivity of 0.9474, Specificity of 0.83332, Balanced Accuracy of 0.8903, Overall Accuracy of 0.9135, FPR of 0.1667, FNR of 0.0526, Precision of 0.9308, F1 0.9390 and MCC of 0.7908. The best parameters obtained for LogReg was C = 10, for KNN was 1300 trees. Similarly, the parameters obtained for Random Forest was max_depth=50, max_features=auto, min_samples_leaf=1, min_samples_split=2 and n_estimators=1000. Similarly, for ET, the best parameters were n_estimators=1500, and Bagging Classifier was built with 1000 trees and Decision Tree as the base classifier. Although Extra Tree performs better than the Random Forest Classifier on False Positive Rate and Specificity, we choose Random Forest Classifier because it performs best on all the other metrics.

### c) Search for the best classifier for time window 26.67 min to 53.34 min

*Table 3: Search for the best classifier for prediction*

| Methods | LogReg | KNN | RF | ET |
|---|---|---|---|---|
| Sensitivity | 0.7755 | 0.9544 | **0.9621** | 0.9452 |
| Specificity | **0.7943** | 0.6369 | 0.6816 | 0.6957 |
| Balanced Accuracy | 0.7849 | 0.7956 | **0.8219** | 0.8205 |
| Overall Accuracy | 0.7796 | 0.8851 | **0.9009** | 0.8907 |
| FPR | **0.2056** | 0.3630 | 0.3183 | 0.3042 |
| FNR | 0.2244 | 0.0455 | **0.0378** | 0.3042 |
| Precision | **0.9310** | 0.9039 | 0.9154 | 0.9175 |
| F1 | 0.8461 | 0.9285 | **0.9324** | 0.6947 |
| MCC | 0.49339 | 0.6434 | **0.6947** | 0.6687 |

LogReg=Logistic Regression, KNN=K Nearest Neighbors, RF=Random Forest, ET = Extra Tree

From **Table 3**, we can see that Random Forest performs the best with a Sensitivity of 0.9621, a specificity of 0.6816, Balanced Accuracy of 0.8219, Overall Accuracy of 0.9009, FPR of 0.3183, FNR of 0.0378, Precision of 0.9154, F1 of 0.9324, and MCC of 0.6947. Note that Logistic Regression has the highest Precision and Sensitivity among all the models tested and has the lowest False Positive Rate. However, we choose Random Forest Classifier because it beats all the other classifiers in other metrics. The best parameters obtained for LogReg was C = 1, for KNN was 1200 trees. Similarly, the parameters obtained for Random Forest was max_depth=40, max_features=auto, min_samples_leaf=2, min_samples_split=2 and n_estimators=800. Similarly, for ET, the best parameters were n_estimators=1000, and Bagging Classifier was built with 1000 trees and Decision Tree as the base classifier.

### d) Search for the best classifier for time window 53.34 min to 80.01 min

*Table 4: Search for the best classifier for prediction.*

| Methods | LogReg | KNN | RF | ET |
|---|---|---|---|---|
| Sensitivity | 0.7699 | 0.9529 | **0.9591** | 0.9437 |
| Specificity | **0.7941** | 0.6144 | 0.6555 | 0.6614 |
| Balanced Accuracy | 0.7820 | 0.7837 | **0.8073** | 0.8020 |
| Overall Accuracy | 0.7750 | 0.8822 | **0.8957** | 0.8847 |
| FPR | **0.2058** | 0.3855 | 0.3444 | 0.3385 |
| FNR | 0.2300 | 0.0470 | **0.0408** | 0.0562 |
| Precision | **0.9340** | 0.9034 | 0.9133 | 0.9134 |
| F1 | 0.8441 | 0.9275 | **0.9356** | 0.6664 |
| MCC | 0.4815 | 0.6206 | **0.6664** | 0.6366 |

LogReg=Logistic Regression, KNN=K Nearest Neighbors, RF=Random Forest, ET = Extra Tree

From **Table 4**, we can see that Random Forest performs the best with a Sensitivity of 0.9591, Specificity of 0.6555, Balanced Accuracy of 0.8073, Overall Accuracy of 0.8957, FPR of 0.3444, FNR of 0.0408, Precision of 0.9133, F1 of 0.9356, and MCC of 0.6664. We can see that Logistic Regression, however, has the highest Specificity and Precision and has the lowest False Positive Rate. However, it does not outperform Random Forest on all the other metrics. Thus, we choose Random Forest as the best classifier. The best parameters obtained for LogReg was C = 0.1, for KNN was 1000 trees. Similarly, the parameters obtained for Random Forest was max_depth=10, max_features='sqrt', min_samples_leaf=2, min_samples_split=2 and n_estimators=200. Similarly, for ET, the best parameters were n_estimators=1000, and Bagging Classifier was built with 1000 trees and Decision Tree as the base classifier.

*e) Searching for the best classifier for time window 80.01 min to 106.58 min*

*Table 5: Search for the best classifier for prediction*

| Methods | LogReg | KNN | RF | ET |
|---|---|---|---|---|
| Sensitivity | 0.7657 | 0.9539 | **0.9699** | 0.9685 |
| Specificity | **0.7915** | 0.6093 | 0.6182 | 0.6203 |
| Balanced Accuracy | 0.7709 | 0.7816 | **0.7941** | 0.7944 |
| Overall Accuracy | 0.7709 | 0.8840 | **0.9181** | 0.9172 |
| FPR | **0.2084** | 0.3906 | 0.3817 | 0.3800 |
| FNR | 0.2343 | 0.0461 | **0.0300** | 0.0315 |
| Precision | 0.9352 | 0.9056 | **0.9363** | 0.9361 |
| F1 | 0.8420 | 0.9291 | **0.9528** | 0.9522 |
| MCC | 0.4706 | 0.6172 | **0.6493** | 0.6462 |

LogReg=Logistic Regression, KNN=K Nearest Neighbors, RF=Random Forest, ET = Extra Tree

From **Table 5**, we can see that Random Forest performs the best with a Sensitivity of 0.9699. Specificity of 0.6182, Balanced Accuracy of 0.7941, Overall Accuracy of 0.9181, FPR of 0.3817, FNR of 0.0300, Precision of 0.9363, F1 0.9528 and MCC of 0.6493. We can see that Logistic Regression, however, has the highest Specificity and has the lowest False Positive Rate. However, it does not outperform Random Forest on all the other metrics. Thus, we choose Random Forest as the best classifier. The best parameters obtained for LogReg was C = 1, for KNN was 1300 trees. Similarly, the parameters obtained for Random Forest was max_depth=20, max_features='sqrt', min_samples_leaf=2, min_samples_split=2 and n_estimators=400. Similarly, for ET, the best parameters were n_estimators=1000, and Bagging Classifier was built with 1200 trees and Decision Tree as the base classifier.

We can see from the results that the performance of the classifiers increases when more features from the Fourier Spectra are included. Moreover, we can see that Logistic Regression, which is a weakly nonlinear model, can predict rogue waves from the normal waves with a considerable degree of accuracy. The results validate the conclusions of Petrova and Soares that weakly nonlinear models are still helpful to predict nonlinear effects [38]. Moreover, the performance of the Tree-based methods like Random Forest and Extra Tree suggests that nonlinear models can capture various types of nonlinearity in the oceans. Likewise, since the Random Forest algorithm is very robust to noise compared to Extra Trees Classifier, it performs the best for all time windows explored in the paper. Moreover, as the prediction time for forecast increases, the Balanced Accuracy also decreases. It suggests that more features are required to forecast rogue waves for longer times.

# 6. CONCLUSIONS

From the results, we can conclude that it is possible to forecast rogue waves with the help of machine learning methods. First of all, we found that Spectral Features are important for forecasting rogue waves. Also, Random Forest outperforms all the other classifiers with Overall Accuracy and the Balanced Accuracy varying from 89.57% to 91.81%, and 79.41% to 89.03%, respectively, depending on the forecast time window.

With the use of model-free evaluation criteria, various Spectral Features, and statistical machine learning methods, the warning time for rogue waves has been improved from the scale of seconds/minutes to a scale of hours. We propose that a similar framework could be used to predict extreme events in other mediums, including but not limited to various nonlinear dispersive media.

# Appendix

This section contains the definitions of the terminologies used in the text.

**Linear Effect/ Nonlinear Effect**: Gaussian Theory of waves assume that the waves can only interact with each other linearly. This is known as linear effects. However, recent experiments have shown that when waves combine, various nonlinear effects may take place. These effects can be studied with the help of nonlinear equations. So, they are also known as nonlinear waves

**Stokes expansion**: In fluid dynamics, a Stokes wave is a non-linear and periodic surface wave on an inviscid fluid layer of constant mean depth. This type of modelling has its origins in the mid-19th century when Sir George Stokes – using a perturbation series approach, now known as the Stokes expansion – obtained approximate solutions for non-linear wave motion

**Steepness**: The wave steepness is defined as the ratio of wave height H to the wavelength λ.

**UKCP**: United Kingdom Climate Projections

**ECMWF**: European Centre for Medium-Range Weather Forecasts

**Pile up**: accumulate

**Wave instability**: There are situations when a wave is capable of extracting energy from the system. It does so by drawing either kinetic energy from pre-existing motion or potential energy from background stratification. In either case the wave amplitude grows over time, and the wave is said to be unstable.

**Nonlinear wave statistics**: The statistics of the wave system when there are nonlinear waves present.

**Modulational Instability**: In the fields of nonlinear optics and fluid dynamics, modulational instability or sideband instability is a phenomenon whereby deviations from a periodic waveform are reinforced by nonlinearity, leading to the generation of spectral-sidebands and the eventual breakup of the waveform into a train of pulses. The phenomenon was first discovered − and modelled − for periodic surface gravity waves (Stokes waves) on deep water by T. Brooke Benjamin and Jim E. Feir, in 1967. Therefore, it is also known as the Benjamin−Feir instability.

**CDIP**: The Coastal Data Information Program (CDIP) specializes in wave measurement, swell modeling and forecasting, and the analysis of coastal environment data.

**Directional Spreading Function**: A directional wave spectrum can be described by $S(f, \theta) = S(f)D(\theta)$ where S(f) is the energy density spectrum and $D(\theta)$ is the directional spreading function at frequency f. A general directional spreading function at frequency f can be expanded in angular Fourier series.

$$D(\theta) = \frac{1}{\pi}(\frac{1}{2} + \sum_{n=1}^{\infty} A_n \cos n\theta + B_n \sin n\theta)$$

Where $A_n$ and $B_n$ are the angular Fourier coefficients.

**Fourier Moments**: The moments derived from the function after Fourier transform. The first moment of a function f(x) is given by

$$\int_{-\infty}^{\infty} x f(x) dx = \frac{-i}{2\pi} \frac{dF(0)}{ds}$$

Likewise, the nth moment is given by

$$\int_{-\infty}^{\infty} x^n f(x) dx = (\frac{-i}{2\pi})^n \frac{d^n F(0)}{ds^n}$$

**KVH criteria**: The method proposed by A.J Kuik, G. Ph van Vledder and L.H. Holthuijsen for routine analysis of pitch-and-roll buoy data. It yields four directional model-free parameters per frequency to provide directional information: the mean direction, the directional width, the skewness, and the kurtosis of the directional energy distribution.

**Minowski distance**: Minowski distance is a generalized metric distance. When $\lambda = 1$, it becomes Euclidean disatance. Similarly, Chebyshev distance is a special case of Minowski distance with $\lambda = \infty$. It can be used for both ordinal and quantitative variables. The formula is

$$d_{ij} = \sqrt[\lambda]{\sum_{k=1}^{n} |x_{ik} - x_{jk}|^\lambda}$$

Where $d_{ij}$ refers to distance between two points and $x_{ik}$ and $x_{jk}$ refer to the coordinates of points i and j.